# Fuzzy Cluster Analysis: Application to Determining Metallicities for Very Metal-poor Stars

Haining Li
Key Lab of Optical Astronomy, National Astronomical Observatories, Chinese Academy of Sciences A20 Datun Road, Chaoyang, Beijing 100101, People's Republic of China; lhn@nao.cas.cn

## Abstract

This work presents a first attempt to apply fuzzy cluster analysis (FCA) to analyzing stellar spectra. FCA is adopted to categorize line indices measured from LAMOST low-resolution spectra, and automatically remove the least metallicity-sensitive indices. The FCA-processed indices are then transferred to the artificial neural network (ANN) to derive metallicities for 147 very metal-poor (VMP) stars that have been analyzed by high-resolution spectroscopy. The FCA-ANN method could derive robust metallicities for VMP stars, with a precision of ∼0.2 dex compared with high-resolution analysis. The recommended FCA threshold value $\lambda$ for this test is between 0.9965 and 0.9975. After reducing the dimension of the line indices through FCA, the derived metallicities are still robust, with no loss of accuracy, and the FCA-ANN method performs stably for different spectral quality from [Fe/H] ∼ −1.8 down to −3.5. Compared with traditional classification methods, FCA considers ambiguity in groupings and noncontinuity of data, and is thus more suitable for observational data analysis. Though this early test uses FCA to analyze low-resolution spectra, and feeds the input to the ANN method to derive metallicities, FCA should be able to, in the large data era, also analyze slitless spectroscopy and multiband photometry, and prepare the input for methods not limited to ANN, in the field of stellar physics for other studies, e.g., stellar classification, identification of peculiar objects. The literature-collected high-resolution sample can help improve pipelines to derive stellar metallicities, and systematic offsets in metallicities for VMP stars for three published LAMOST catalogs have been discussed.

*Unified Astronomy Thesaurus concepts:* Population II stars (1284)

*Supporting material:* machine-readable table

## 1. Introduction

In the past decade, large-scale spectroscopic survey projects have resulted in huge amounts of observational data for stars in the Milky Way, e.g., the Sloan Extension for Galactic Understanding and Exploration (SDSS/SEGUE; Yanny et al. 2009), the Apache Point Observatory Galactic Evolution Experiment (SDSS/APOGEE; Majewski et al. 2017), the Large Sky Area Multi-Object Fiber Spectroscopic Telescope survey[1] (LAMOST; Zhao et al. 2006; Cui et al. 2012), the Radial Velocity Experiment (RAVE; Steinmetz et al. 2006), the Gaia-ESO Public Spectroscopic Survey (GES; Gilmore et al. 2012), and the GALAH Survey (De Silva et al. 2015). Such unprecedentedly large databases of stellar spectra provide great opportunities to explore the formation and evolution of our Galaxy through studying various types/populations of stars. Among these, metal-poor stars have been one of the major targets, as they are of crucial importance to the field of Galactic archeology (Frebel & Norris 2015). Very metal-poor (VMP) stars with [Fe/H] < −2.0[2] provide the fossil record of the early chemical history of the Galaxy and early generations of stars (Beers & Christlieb 2005). In particular, statistical studies of VMP stars provide crucial constraints on chemical evolution models of the Milky Way (see, e.g., Bromm & Yoshida 2011; Nomoto et al. 2013), and important evidence of ancient accretion (e.g., Venn et al. 2004; Matsuno et al. 2019).

Current sources of VMP star candidates include selections from objective prism surveys, such as the HK Survey (Beers et al. 1992) and Hamburg/ESO Survey (Christlieb et al. 2008), low-resolution spectroscopic surveys, including SEGUE (Yanny et al. 2009) and LAMOST (Zhao et al. 2012), and photometric survey projects including the SkyMapper survey (Keller et al. 2007) and the Pristine survey (Starkenburg et al. 2017). Thanks to these efforts, the sample size of candidate VMP stars has been remarkably increased (e.g., Frebel et al. 2006; Schlaufman & Casey 2014; Li et al. 2018b). A number of important discoveries have been made, based on high-resolution follow-up observations of these candidates, e.g., stars with unprecedentedly low iron abundance (Keller et al. 2014) or total metal content (Caffau et al. 2011) and possible signatures of supermassive pair-instability supernovae (Aoki et al. 2014). And, after the launching of Gaia, specifically the release of DR2 (Gaia Collaboration et al. 2018), statistical studies on kinematics of the VMP stars became possible and have already been proven to be an invaluable source to trace the formation history of the Milky Way (e.g., Sestito et al. 2019, 2020; Yuan et al. 2020). Therefore, searching for more VMP stars in current and future large-scale survey projects is very important to deepening our understanding of early history of the Milky Way and local universe.

---

[1] http://www.lamost.org/public
[2] $[A/B] = \log(N_A/N_B)_\star - \log(N_A/N_B)_\odot$, where $N_A$ and $N_B$ are the number densities of elements A and B, respectively, and ⋆ and ⊙ refer to the star and the Sun, respectively.

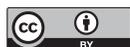

In the past years, quite a number of new methods and pipelines have been designed to derive stellar parameters including metallicities for large database of stellar spectra obtained from survey projects like LAMOST, SDSS/SEGUE, APOGEE (Majewski et al. 2017), etc. For example, the SEGUE Stellar Parameter Pipeline (SSPP; Lee et al. 2008) adopts a combination of different methods, including template matching with synthetic spectra and the neural network approaches; ULySS (Koleva et al. 2009) fits spectroscopic observations against a linear combination of nonlinear model components convolved with a parametric line-of-sight velocity distribution, and has been adopted by the official LAMOST Stellar Parameters Pipeline (LASP; Luo et al. 2015); the Cannon (Ness et al. 2015) provides a good example of a data-driven approach to derive stellar labels for a large number of spectra; the Payne (Ting et al. 2019) leverages the full usage of physical spectral models and fits all labels simultaneously. All these methods are proven to be capable of deriving stellar parameters for a large number of stars based on stellar spectra; however, in most cases, due to a limited number of training sample objects or insufficiently wide coverage in the template space in the low-metallicity region, most pipelines cannot predict robust metallicities for VMP stars. One of the exceptions would be SSPP, which has used the KP line index that measures the strength of the Ca II K line to calibrate stellar metallicities down to [Fe/H] $\sim -4.0$. Therefore, determination of stellar parameters, especially metallicities for VMP stars, requires specific treatment or analysis, especially for large-scale survey projects covering a wide sky area and various types of stellar objects that cover a wide range of stellar parameters.

One of the characteristics of VMP stars is that their spectra do not include as many absorption features as metal-rich stars, which makes it relatively difficult to extract sufficient information that is useful to identify VMP stars. Efforts have been carried out in the past decades to identify the most metallicity-sensitive spectral features for categorizing VMP stars (e.g., Beers et al. 1999), and a few atomic lines, including the Ca II HK lines at 3933 Å and the Ca triplet at 8662 Å, have been found to be quite sensitive to metallicities even in very low-metallicity regions (Christlieb et al. 2008; Xu et al. 2013). The strong features of Ca II HK lines have been widely used to identify metal-poor stars, not only in spectroscopic data (e.g., the KP line index used in SSPP), but also utilized in the design of photometric filters (e.g., the v-filter of the SkyMapper Survey, the Ca HK filters of the Pristine Survey). It is, of course, a way to derive reliable metallicities for VMP stars relying on such features that have been proven to be efficient based on previous experiences, but it is not very convenient when the observed spectra do not contain features that are well studied (e.g., the Ca II K line). Therefore, it would be interesting and also helpful to investigate the possibility of establishing more objective methods to identify sensitive/ efficient (combination of) spectral features to derive robust metallicities for VMP stars. And that is why this paper presents an effort as a pioneering test of applying a mathematical clustering analysis method—fuzzy cluster analysis (FCA) on astronomical data.

In mathematics, "fuzzy" means "vague," and as a new subject developed after classical mathematics and statistical mathematics, fuzzy mathematics is a set of mathematical methods to describe data sets where the membership of elements is not binary. In production practice, scientific experiments, and daily life, people often encounter ambiguous (i.e., fuzzy) concepts or phenomena, e.g., classifying objects as light or heavy, dynamic or static, that have some ambiguity or "fuzziness" involved. Therefore, when they need to be analyzed quantitatively, the tool of fuzzy mathematics is preferred. Fuzzy clustering is a mathematical method that classifies objects using fuzzy mathematical language, which is a group of methods associated with describing sets where the membership is probabilistic. Its basic rule is to divide the data set into several classes or clusters, in accordance with the principle of "minimize the similarity between classes and maximize the similarity within classes"; the difference between samples from different classes should be as large as possible, while the difference between samples within each class should be as small as possible. Through fuzzy clustering, each sample will be assigned to a category with a certain degree of membership/probability. Based on such principle, the method of fuzzy cluster analysis (FCA) has been proposed. In the past 40 years, regarded as a new, practical, and effective mathematical classification method, FCA has been widely used in many fields, such as pattern recognition, data mining, and fuzzy control, and has obtained excellent results in the fields of geoscience and hydrology (Rao & Srinivas 2006; Langer et al. 2011; Chen et al. 2015), but has not yet been applied much to astronomical data.

Among all classification methods, $K$—means clustering has been the most commonly used in analyzing astronomical data, e.g., searching for extremely metal-poor galaxies (Sánchez Almeida & Allende Prieto 2013), classification of stellar spectra obtained by SDSS/SEGUE and APOGEE (Sánchez Almeida et al. 2016; Garcia-Dias et al. 2018). Classical methods including $K$—means belong to "hard classification," which adopts a nonprobabilistic model and deduces the classification result from the decision function, where the sample is definitely assigned to one class. The "hard classification" methods are more suitable to work on continuous data (or continuous variables) that can be arbitrarily valued in a certain interval. This is obviously not true for observational or experimental data in reality, in which a discrete set of points with uncertainty are observed and need to be classified. The FCA method belongs to the other type, "soft classification," where the probability that a sample belongs to each class corresponds to a membership vector, and the classification returns the one with the highest probability. In real life, actual measurements including observational data in astronomy are not continuous variables, which involves fuzziness/ambiguity in grouping. In other words, the traditional hard classification algorithms do not consider the fuzziness of data, while soft classification is more in line with the actual situation. Compared with requiring precise numbers and assuming precise relations in classical clustering methods, fuzziness and imprecision of the data are considered with the soft classification algorithms. The fuzzy clustering method can also be regarded as a theoretical method for data compression, in which a large sample can be converted into a small number of representative prototypes or similar groups using fuzzy mathematics theory. Therefore, soft classifications such as the FCA method provide more flexible and applicable conditions for processing more information, and are obviously worth trying to apply to astronomical data analysis, especially in this new age of big data.

Table 1
Definition of Line Indices

| Index (1) | Line Band (Å) (2) | Blue Band (Å) (3) | Red Band (Å) (4) |
|---|---|---|---|
| CaHK | 3900.000−4000.000 | 3837.000−3877.000 | 4040.000−4080.000 |
| K24 | 3921.700−3945.700 | 3903.000−3923.000 | 4000.000−4020.000 |
| K12 | 3927.700−3939.700 | 3903.000−3923.000 | 4000.000−4020.000 |
| K6 | 3930.700−3936.700 | 3903.000−3923.000 | 4000.000−4020.000 |
| He I4026 | 4020.200−4032.200 | 4000.000−4020.000 | 4144.000−4164.000 |
| HD24 | 4089.800−4113.800 | 4000.000−4020.000 | 4144.000−4164.000 |
| HD12 | 4095.800−4107.800 | 4000.000−4020.000 | 4144.000−4164.000 |
| CN1 | 4142.125−4177.125 | 4080.125−4117.625 | 4244.125−4284.125 |
| CN2 | 4142.125−4177.125 | 4083.875−4096.375 | 4244.125−4284.125 |
| Ca4227 | 4222.250−4234.750 | 4211.000−4219.750 | 4241.000−4251.000 |
| G4300 | 4281.375−4316.375 | 4266.375−4282.625 | 4318.875−4335.125 |
| G1 | 4297.500−4312.500 | 4247.000−4267.000 | 4362.000−4372.000 |
| Fe4383 | 4369.125−4420.375 | 4359.125−4370.375 | 4442.875−4455.375 |
| Ca4455 | 4452.125−4474.625 | 4445.875−4454.625 | 4477.125−4492.125 |
| He I4471 | 4465.700−4477.700 | 4415.000−4435.000 | 4490.000−4510.000 |
| Fe4531 | 4514.250−4559.250 | 4504.250−4514.250 | 4560.500−4579.250 |
| Hbeta | 4847.875−4876.625 | 4827.875−4847.875 | 4876.625−4891.625 |
| Fe5015 | 4977.750−5054.000 | 4946.500−4977.750 | 5054.000−5065.250 |
| Mg1 | 5069.125−5134.125 | 4895.125−4957.625 | 5301.125−5366.125 |
| Mg2 | 5154.125−5196.625 | 4895.125−4957.625 | 5301.125−5366.125 |
| Mgb | 5147.125−5192.625 | 5142.625−5161.375 | 5191.375−5206.375 |
| Fe5270 | 5245.650−5285.650 | 5233.150−5248.150 | 5285.650−5318.150 |
| Fe5335 | 5312.125−5352.125 | 5304.625−5315.875 | 5353.375−5363.375 |
| Fe5406 | 5387.500−5415.000 | 5376.250−5387.500 | 5415.000−5425.000 |
| Fe5709 | 5696.625−5720.375 | 5672.875−5696.625 | 5722.875−5736.625 |
| Fe5782 | 5776.625−5796.625 | 5765.375−5775.375 | 5797.875−5811.625 |
| NaD | 5876.875−5909.375 | 5860.625−5875.625 | 5922.125−5948.125 |

This paper is the first attempt to apply the mathematical fuzzy clustering method to analyzing stellar spectra, specifically speaking, to test the feasibility of using FCA method to group observational data (line indices) and to retain the most relevant information (by reducing the dimension) of the input to estimate metallicities for VMP stars based on LAMOST low-resolution spectra. It would be helpful to investigate the efficiency of fuzzy clustering and its effect on the derived parameters. Section 2 describes the methods and procedures that we adopt to derive stellar metallicities from LAMOST low-resolution spectra. Section 3 demonstrates the validity of this approach through comparison with high-resolution analysis and other methods and discusses the general clustering results such as the implication of FCA classification of line indices. Section 4 briefly summarizes the results of this work, together with the potential of this method for future work in the era of large scale surveys.

## 2. Data and Methodology

The analysis of this paper is based on spectra obtained through LAMOST DR5, which was released in 2019 July. Since we are interested in stars with very little metal that only present relatively weak absorption lines, and the precision of metallicities derived from low-resolution spectra for VMP stars are clearly relevant to the spectra quality (as shown by Li et al. 2018b), we started by selecting all stellar spectra with signal-to-noise ratios (S/N) higher than 20 in both the $g$ and $r$ bands. Furthermore, as the purpose of this work is to cluster and identify the most efficient features for VMP stars, it is also important that the analysis is indeed performed on "true" VMP stars, and hence the selected LAMOST targets have been cross-matched with VMP stars confirmed by high-resolution spectroscopy from SAGA database (Suda et al. 2011) and JINAbase (Abohalima & Frebel 2018). The procedure has thus resulted in a VMP list including 147 VMP stars, as listed in Table 3 in the appendix.

### 2.1. Measured Line Indices

The analysis is performed on line indices. Therefore we have adopted the same list of line indices as Li et al. (2018b), which includes 27 line indices based on Lick indices[3] and also the line indices used for the SSPP. Definitions of the measured line indices are listed in Table 1. Corresponding line indices were computed for all the program stars.

### 2.2. Fuzzy Cluster Analysis

The fuzzy clustering method constructs a fuzzy matrix (as defined in Appendix A) according to the properties of the research object. Then, the fuzzy clustering method determines the clustering relation according to a membership degree/probability that is determined by the "fuzzy relation" (also defined in Appendix A between samples). In such a way, fuzzy clustering is able to divide the data set into multiple classes or clusters, so that the data difference between each class should be as large as possible and the difference between each data element in one class should be as small as possible. In other words, the principle is to minimize the similarity between classes while maximizing the similarity within classes. The basis of relevant definition and theory of fuzzy clustering is described in detail in Appendix A, and here, we briefly introduce the procedure that has been adopted for the fuzzy cluster analysis (FCA) on the measured line indices of our program stars.

---

[3] http://astro.wsu.edu/worthey/html/index.table.html

### 2.2.1. Data Standardization

In practical problems, different data may have different units, and in order for data with different units to be compared, the data need to be transformed appropriately, which is also referred to as "data standardization." In order to make the original data fit the requirements of fuzzy clustering, the original data matrix A must be standardized, which means that it must be transformed into fuzzy matrix $\tilde{A}$ by appropriate data transformation. Generally, according to the requirements of the fuzzy matrix, the data is compressed into the interval [0, 1].

Suppose domain $U = u_1, u_2,...,u_n$ is the classified object (in this case the 147 selected program stars), and each object is represented by $m$ factors (the 27 line indices measured for each star), that is: $u_i = \{x_{i1}, x_{i2},\cdots,x_{im}\} (i = 1, 2,\cdots,n)$ (where $n = 147$ and $m = 27$). We can then get the original data matrix $\tilde{A} = (x_{ij})_{(n \times m)}$. The element of the matrix can be defined by translation, the transformation of the standard deviation

$$x_{ik}^a = \frac{x_{ik} - \bar{x}_k}{s_k}, (i = 1, 2,...,n; k = 1, 2,...,m), \quad (1)$$

where $\bar{x}_k = (1/n) \sum_{i=1}^{n} x_{ik}$, and $s_k = \sqrt{(1/(n-1)) \sum_{i=1}^{n} (x_{ik} - \bar{x}_k)^2}$, $(k = 1, 2,...,m)$.

If, after the above translation, there are still some $x_{ik}^a$ not belonging to the [0, 1] interval, namely $x_{ik}^a \notin [0, 1]$, it will carry to the next step of translation, range transformation, defined as:

$$x_{ik}^c = \frac{x_{ik}^a - \min\{x_{ik}^a\}}{\max\{x_{ik}^a\} - \min\{x_{ik}^a\}}, (i, k = 1, 2,...,n). \quad (2)$$

After the transformation of the original measurement data, we can get $0 \leqslant x_{ik}^c \leqslant 1$, where the absolute value of each measured result can be changed within the interval [0, 1] and no longer has unit (dimension), or, in other words, the influence of data units in the calculation and analysis process is eliminated. Therefore the original data matrix can be obtained through $\tilde{A} = (x_{ik}^c)_{n \times m}$, which transforms the observational data into a more convenient form for the following fuzzy cluster analysis.

### 2.2.2. Membership Function and Closeness Degree

The next step is also known as calibration, which is to mark the statistic $r$ that measures the similarity between the classified objects $r_{ij} (i, j = 1, 2,...,n)$. Assuming the domain as $U = u_1, u_2,...,u_n$, where each element is a sample, we can then establish the similarity relation $\tilde{R}$ on $U$, where $\tilde{R}$ represents the fuzzy similarity matrix $r_{ij} = \tilde{R}(u_i, u_j)$, and $r_{ij}$ is also referred to the similarity coefficient. Each sample then becomes an $m$ −dimensional vector, that is: $u_i = x_{i1}, x_{i2},...,x_{im}$ ($i = 1, 2,..., n$). And the following formula can be used to determine the similarity coefficient $r_{ij}$:

$$r_{ij} = \frac{\sum_{k=1}^{m}(x_{ik} \wedge x_{jk})}{\sum_{k=1}^{m}(x_{ik} \vee x_{jk})}, (x_{ij} > 0; \quad i, j = 1, 2,...,m), \quad (3)$$

where $\wedge$ represents the minimum operation, and $\vee$ represents the maximum operation.

In order to accurately determine the fuzzy similarity matrix, the concepts of "membership function," "fuzzy set," and "lattice closeness degree" have to be defined first. We use $\mu_{\tilde{A}(x)}$ to describe the degree to which $x$ belongs to the data set $\tilde{A}$, which is called the membership function of the data set $\tilde{A}$. The closer $\mu_{\tilde{A}(x)}$ is to 1, the more likely $x$ belongs to $\tilde{A}$. Suppose $\Omega$ refers to our sample space composed of measured line indices and $x$ being an element of $\Omega$, if $\tilde{A}(x)$ is in the interval [0,1] corresponding to any element $x$ of $\Omega$, $\tilde{A}$ is called a fuzzy set over $\Omega$. Therefore, the mathematical expression of the fuzzy set is given by:

$$\tilde{A} = \{(x, \mu_{\tilde{A}}(x)) | x \in \Omega\}. \quad (4)$$

For any arbitrary $x \in \Omega$, the mapping can be given as follows:

$$\Omega \to [1, 0] \quad x | \to \tilde{A}(x) \in [1, 0]. \quad (5)$$

Accordingly, the set defined as $\tilde{A} = \{(x | \tilde{A}(x))\}$ is referred to the "ordered pair," and $\forall x \in \Omega$ is the subset contained by $\Omega$, which is referred to as a fuzzy set.

Regarding the subject investigated in this paper, the set denoted by $\tilde{A}$ is referred to the "associated set," whose elements are the influencing factors, for this case the line indices related to the physical parameters provided from observational data: in this case, the line indices measured for the program stars. Thus, clearly, $\tilde{A}$ is a fuzzy subset of the set $\Omega$ (where $\Omega$ is a sample space).

Now the membership function for the metallicity-determination problem can be assigned by that of the influencing factors to the fuzzy subset $\tilde{A}_j (j = 1, 2,...,m)$, where $m = 27$ (corresponding to the 27 line indices measured for each program star). A corresponding membership function can then be established based on the fuzzy mathematics theory, adopting the following equations:

$$\mu_{\tilde{A}_j}(x) = \exp\left[-\left(\frac{x - a_j}{b_j}\right)^2\right], \quad (6)$$

$$a_j = \frac{1}{n}\sum_{i=1}^{n} x_{ij}, \quad (7)$$

$$b_j = \sqrt{\frac{1}{n-1}\sum_{i=1}^{n}(x_{ij} - a_j)^2}, \quad (8)$$

where $a_j$ is the mean value, $b_j$ is the sample standard deviation, and $n = 147$, which refers to the number of program stars. In statistics, the average sample difference is divided by the degree of freedom $(n - 1)$, referring to the degree of free choice.

The fuzzy closeness degree is actually to judge the degree of closeness between two fuzzy sets. Suppose $\delta(\tilde{A}, \tilde{B})$ is the fuzzy closeness degree between $\tilde{A}$ and $\tilde{B}$ and satisfies $0 \leqslant \delta(\tilde{A}, \tilde{B}) \leqslant 1$. The larger $\delta(\tilde{A}, \tilde{B})$ is, the closer the two fuzzy subsets are, and the smaller $\delta(\tilde{A}, \tilde{B})$ is, the more distant the two fuzzy subsets become. In order to quantitatively define the fuzzy closeness degree, the concepts of inner product and outer product need to be introduced. Let $\tilde{A}$ and $\tilde{B}$ be two fuzzy subsets on the domain $\Omega = \{u_1, u_2,...,u_n\}$, then the inner product and outer product of $\tilde{A}$ and $\tilde{B}$ can be respectively defined as follows:

$$\tilde{A} \bullet \tilde{B} = V_{i=1}^{n}(\mu_{\tilde{A}}(u_i) \wedge \mu_{\tilde{B}}(u_i)), \quad (9)$$

$$\tilde{A} \otimes \tilde{B} = \Lambda_{i=1}^{n}(\mu_{\tilde{A}}(u_i) \vee \mu_{\tilde{B}}(u_i)). \quad (10)$$

In fuzzy set theory, the larger the inner product is, the closer the fuzzy set will be, while the larger the outer product is, the more distant the fuzzy set will be. Therefore, by combining the inner product and the outer product, the lattice closeness degree between the two fuzzy subsets can be defined as:

$$\delta(\tilde{A}, \tilde{B}) = \frac{1}{2}[\tilde{A} \bullet \tilde{B} + (1 - \tilde{A} \otimes \tilde{B})]. \quad (11)$$

*2.2.3. Fuzzy Similarity Matrix and Clustering*

The next step is to establish the fuzzy similarity matrix $\tilde{R} = (r_{ij})_{m \times m}$ using the above described lattice closeness degree. For the problem studied in this paper, the concrete expression of the fuzzy similarity matrix is as follows:

$$r_{ij} = \exp\left[-\left(\frac{a_j - a_i}{b_i + b_j}\right)^2\right] (i, j = 1, 2, \ldots, m), \quad (12)$$

where $m$ is the so-called number of influencing factors, the number of line indices measured for each program star, while $a_i$ and $a_j$ respectively refer to the mean value of the $i^{th}$ and $j^{th}$ influencing factor, as defined in Equation (7), and $b_i$ and $b_j$, respectively, refer to the standard deviation of the corresponding influencing factors, as defined in Equation (8).

Commonly used fuzzy clustering methods include the Boolean matrix method, direct clustering, and the transitive closure method, which has been adopted for this study. According to the fuzzy mathematical theory, the established fuzzy matrix $\tilde{R}$ usually has reflexivity and symmetry rather than transitivity (for the definition of these mathematical properties, readers may refer to Appendix A). However, in the following process of calculation and analysis for FCA, clustering can only be performed on a fuzzy equivalent, matrix into which we have to transform the fuzzy similarity matrix $\tilde{R}$.

For $\tilde{R}$, there is a minimum natural number $k$ ($k \leqq n$), and $\tilde{R}^k$ is the corresponding fuzzy equivalent matrix, where for all natural numbers $c$ greater than $k$, there is always $\tilde{R}^c = \tilde{R}^k$. The method of the transformation is to find the transitive closure $t(\tilde{R})$ of $\tilde{R}$ by using the square method, which is a successive square multiplication of similar matrices in the calculation (as explained below), so that $t(\tilde{R}) = \tilde{R}^0$.

In fuzzy clustering analysis, to transform similar matrices into equivalent matrices, successive square multiplication of similar matrices is needed in the calculation process. In fact, each time of calculation is a matrix square multiplication operation until the transitive closure is determined. The transitive closure of similarity relation $R$ is calculated based on the following expression:

$$\tilde{R} \to \tilde{R}^2 \to \tilde{R}^{2^2} \to \cdots \to \tilde{R}^{2^k} = \tilde{R}^{2^{k+1}}. \quad (13)$$

Because it is a successive square, this algorithm is called the "square method."

After that, the $n$ order fuzzy similarity matrix $\tilde{R}$ has been transformed into the $n$ order fuzzy equivalent matrix $\tilde{R}^0$. Since the derived fuzzy relation is a fuzzy equivalence relation, for any $\lambda \in [0, 1]$, the truncated $\lambda$−intercept matrix is also an equivalence relation. Based on this, for any given threshold value of $\lambda \in [0, 1]$, a corresponding ordinary equivalence relation $\tilde{R}_{\lambda}^0$ can be obtained, and then the $\lambda$ horizontal classification can be determined.

Once the transitive closure of the fuzzy matrix $\tilde{R}$ has been derived, we can start to choose the criteria for deleting data columns, which in this study becomes: how can we establish the proper criteria that can, at the same time, reduce the dimension of the observed line indices while retaining sufficient information to derive reliable metallicities for the program stars? In order to do that, the criterion has been determined in such a way that it should maximize the information under the condition that the number of measured line indices to be removed is determined. For example, there are 27 line indices measured for each of our program stars, which can be divided into 15 categories as a result of classification returned from the FCA when adopting a threshold value of $\lambda = 0.9990$, among which nine categories contain only one line index, while the other six categories contain more than two indices. Therefore, it is possible to remove (at least) one line index from each of these six categories when the selected line index satisfies the following condition:

$$\text{Min } err = \sum_{i=1}^{n-k}(\bar{d}_{ik} - \bar{d}_i)^2, \quad (14)$$

where $\bar{d}_i$ is the average value of row $i$, and $\bar{d}_{ik}$ is the average value of row $i$ after column $k$ is deleted, while $n$ is the total number of measured samples, and $k$ is the number of deleted columns.

*2.3. ANN Analysis on the FCA Grouped Data*

So far, we have figured out how to categorize the measured line indices of the 147 program stars using the FCA method. The grouped and dimension-reduced line indices should be treated as the input to a parameter-estimation procedure, which will be able to test the efficiency of FCA in extracting the most relevant information for estimating metallicities for VMP stars and to further explore the effect on metallicity precision after including FCA into the data analysis. Among all mathematical algorithms, the artificial neural network (ANN) has been quite widely used in analyzing astronomical data, which has been proven to be rather efficient for stellar parameter estimation (e.g., quite successful application as part of SSPP), even compared with more recent algorithms of machine learning (Flores et al. 2021). Therefore, we decide to adopt the ANN approach for this test, using the classical back-propagation (BP) network.

The BP network is a multilayer feed-forward neural network trained according to the back-propagation algorithm of error. Through the training of sample data, the network weight and threshold value are constantly modified to make the error function descend along the direction of negative gradient and approximate the expected output. It is a widely used neural network model, which is mainly used for function approximation, model recognition and classification, data compression, and time series prediction. The BP network is composed of an input layer, a hidden layer, and an output layer. The hidden layer can have one or more layers. In this study, the three-layer BP network model is selected. The $s$-type tangent function is selected as the excitation function of hidden layer neurons $f(x) = 1/(1 + \exp(-x))$. The network weight and threshold value are continuously adjusted by the back error function minimize $E = (\sum_i (t_i + o_i)^2)/2$, where $t_i$ is the expected output value and $o_i$ is the calculated output value of the network.

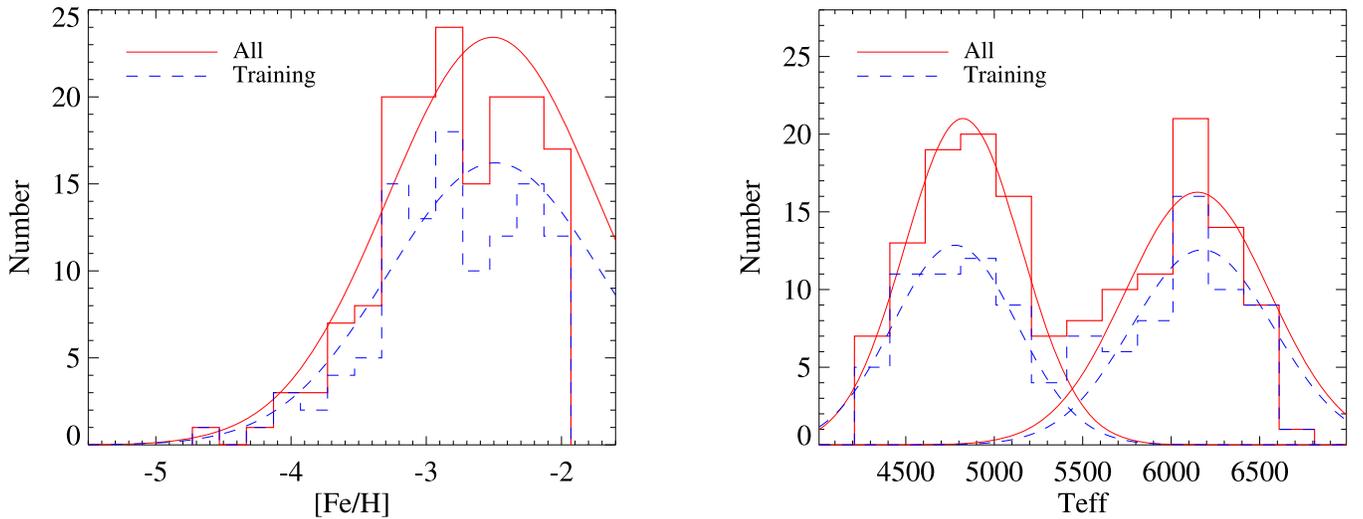

**Figure 1.** Comparison between the distribution of the training sample (red) and the whole LAMOST-SAGA VMP sample (blue) for metallicities (left) and temperature (right).

In the practical calculation, the first step is to design the network structure. Previous studies show that a neural network with a hidden layer can approximate a nonlinear function with arbitrary precision as long as there are enough hidden nodes. Therefore, we have adopted a BP network with three layers of multiple input and single output with one hidden layer to establish a prediction model. After that, a proper excitation function needs to be selected. BP neural network usually adopts Sigmoid differentiable function and linear function as the excitation function of the network. In this paper, the s-type tangent function tansig is selected as the excitation function of hidden layer neurons. As the output of the network is normalized to the range of [−1, 1], the prediction model selects the s-type logarithmic function logsig as the excitation function of the neurons in the output layer. As for the realization of the BP network forecast, the specific implementation includes two main steps.

The training sample data are used as input for the network after normalization and set the excitation function of the hidden layer and the output layer. Note that the excitation function of hidden layer is the hyperbolic tangent function (tanh) or tansig, which can replace the logsig function. The value range of tansig is [−1, 1], while the tansig function is expressed as follows:

$$\text{tansig}(x) = (e^x - e^x)/(e^x + e^x). \quad (15)$$

The excitation function of the network output layer is the logsig function, which is the sigmoid function in logistic regression, and its value range is [0, 1]. The logsig function is expressed as follows:

$$\text{logsig}(x) = \frac{1}{1 + e^{-x}}. \quad (16)$$

Once the network training function is determined, the network training could start. In this case, the built-in function traingdx (from MATLAB), which is described as a multilayer perceptron with momentum gradient descent and adaptive learning rate back propagation, has been adopted. The network performance function adopts mse (the average squared error performance function). The number of network iterations, expected error, and learning rate are set to be 3000, 0.000001, and 0.01, respectively.

### 2.4. Application to the Observational Data

Based on the above description, a procedure of deriving metallicities from low-resolution spectra through FCA-ANN has been established, where for the first and most important step, the fuzzy clustering method was used to classify all measured line indices. After classification, for each category (with more than one line index), the least relevant line index was removed before forwarding to the next step as prediction data. And then the ANN method was used to predict the metallicities based on the clustered and dimension-decreased grid of line indices.

The metallicity [Fe/H] was estimated for all 147 selected VMP stars through the FCA-ANN method, as described above and based on the set of 27 line indices measured through LAMOST spectra. Since the FCA-ANN analysis was performed on a relatively small sample, a proportion of 7:3 between the training and testing data sets was adopted. Targets for the training set were randomly selected from each metallicity bin to reach a fraction of about 70%, and as seen from Figure 1, for the key stellar parameter for this work, [Fe/H], the distribution of the training set can well represent the whole sample. Please note that, since there was a very limited number of stars with [Fe/H] < −4.3, all objects within this metallicity range were kept in the training set to make sure that it covers the required metallicity coverage. Besides, we have also checked the distribution of temperature, which could be important as a basic stellar parameter, and from the right panel of Figure 1, it is shown that the training set also shares similar distribution in $T_{\text{eff}}$.

## 3. Discussion

### 3.1. FCA Clustering of Line Indices

The key parameter to control the clustering through FCA is the threshold value $\lambda$, and by default, it could be any number between 0 and 1. The larger the value, the more groups will be resulted from the clustering. Before deciding which threshold value(s) should be adopted, one should keep in mind that there

**Table 2**
Clustering of Line Indices with Different Thresholds

| (1) | 0.9865 (2) | 0.9885 (3) | 0.9965[a] (4) | 0.9985 (5) | 0.9990 (6) |
|---|---|---|---|---|---|
| CaHK | G1 | G1 | G1 | G1 | G1 |
| K24 | G2x | G2x | G2 | G2 | G2 |
| K12 | G2 | G2 | G3 | G3 | G3 |
| K6 | G3 | G3 | G4x | G4x | G4x |
| He I4026 | G4 | G4 | G5 | G5 | G5 |
| HD24 | G3 | G3 | G6 | G6 | G6 |
| HD12 | G3x | G3x | G7 | G7 | G7 |
| CN1 | G5x | G5x | G8 | G8 | G8 |
| CN2 | G5 | G5 | G8 | G8 | G9 |
| Ca4227 | G6 | G6 | G9x | G9x | G10x |
| G4300 | G3 | G3 | G4 | G4 | G4 |
| G1 | G3 | G3 | G6x | G6x | G6x |
| Fe4383 | G5 | G5 | G8 | G8x | G9x |
| Ca4455 | G6 | G6 | G9 | G9x | G10 |
| He I4471 | G4 | G4 | G5 | G10 | G11 |
| Fe4531 | G5 | G5 | G8x | G8 | G8x |
| H_beta | G3 | G3 | G4 | G11 | G12 |
| Fe5015 | G5 | G5 | G8 | G8 | G9 |
| Mg1 | G5 | G5 | G8 | G8 | G9 |
| Mg2 | G5 | G7 | G10 | G12 | G13 |
| Mgb | G5 | G5 | G8 | G8 | G9 |
| Fe5270 | G5 | G5 | G8 | G13x | G14 |
| Fe5335 | G5 | G5 | G8 | G13 | G15 |
| Fe5406 | G6x | G6x | G9 | G9 | G10 |
| Fe5709 | G4x | G4x | G5 | G5 | G5 |
| Fe5782 | G4 | G4 | G5x | G5x | G5x |
| NaD | G5 | G5 | G8 | G8 | G9 |

**Notes.** Columns (2) through (6) present the clustering results of all measured line indices when adopting different threshold values for the FCA. G1, G2, etc. refer to the categorized groups of indices, and any group ID ended with an "x" (e.g., G4x, G5x) means that this line index has been excluded automatically from corresponding group for the metallicity estimation.
[a] adopting a threshold of 0.9975 leads to the same clustering results.

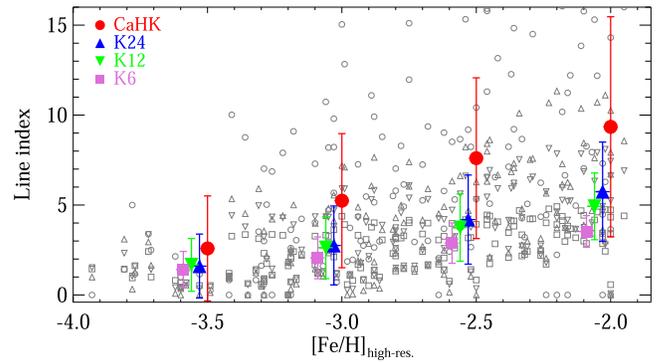

**Figure 2.** Correlation between metallicities and the four Ca II HK-line-focused indices (CaHK, K24, K12, and K6). Individual line index measurements are shown in hollow symbols, with circles, triangles, upside-down triangles, and squares, respectively, for CaHK, K24, K12, and K6. Average values in different metallicity bins are presented in filled symbols, in red, blue, green, and magenta, respectively, for CaHK, K24, K12, and K6.

is always a reasonable range to set the threshold value $\lambda$, since values that are too large (too many groups) will result in a large number of groups with a single member, and thus cannot be helpful to further reduce the dimension of input or efficiently extract the most relevant information. On the other hand, values that are too small (too few groups) will not be able to provide sufficient information to derive reliable metallicities. For example, in our case of using the set of 27 line indices to estimate stellar metallicities, adopting a threshold value of 0.9990 may lead to 15 groups, while adopting 0.9865 will result in 6 groups. And for any group that has more than one indices, it is then possible to remove (at least) one of the group member index, reducing the dimension of input, before transferring the line indices to the next step (e.g., ANN) for metallicity estimations.

A number of different threshold values have been tested, and a few examples are shown in Table 2. Adopting threshold values of 0.9865, 0.9885, 0.9965 (or 0.9975), 0.9985, and 0.9990 will result in 6, 7, 10, 13, and 15 groups, respectively. If the least relevant index (in its group) is automatically excluded whenever a group has more than one indices, each choice of threshold value would be able to reduce the dimension of input indices by 5, 5, 5, 6, and 6, respectively (all excluded indices are marked out by adding an "x" to the end of its group ID). And, for current the sample of VMP stars, the preferred threshold value is about $\lambda = 0.9965 \sim 0.9975$, which will be discussed through the following text.

As shown in Table 2, when adopting different threshold values, the FCA would return different combinations of grouping. The measured line indices are assigned to different groups, though the FCA based on different information or components that are embedded in the observed spectrum: each group shall represent certain features of the observed spectra, which should be relevant to physical parameters (not only limited to metallicities). After that, the FCA would further estimate the sensitivities/relations to metallicities for indices in each group, which would then become the basis of index exclusion in corresponding groups. However, in some cases, it seems that line indices that measure similar spectral features are grouped into different categories. Take the Ca II HK-line-focused indices (K6, K12, K24, CaHK) as an example. If we check the relation between their strengths and the metallicity (as shown in Figure 2), it is clear that the four indices present different strengths (and may cover different information) at the same metallicity. For example, the CaHK exhibits the steepest/largest slope, and K12 and K24 are following similar trends, while the slope of K6 is not as steep as the above three indices. This is not unexpected, since the four line indices correspond to different line bands (as defined in Table 1), where CaHK covers the widest range and thus exhibits the largest index value, while K6 measures the smallest range and thus exhibits the smallest value. It should be noticed that the strengths of the four indices become more similar at extremely low-metallicity range: the difference in strength for K6, K12, and K24 become rather negligible at [Fe/H] < −3.5. This is also why a KP index has been designed to convert among the three line indices according to the relative strength of the Ca II K line, which has been adopted to estimate the metallicity for metal-poor stars, e.g., by the SSPP, the HES pipeline. However, since the number of extremely metal-poor stars in our sample is too small, e.g., only 14 objects have [Fe/H] < −3.5, the K6 index is not assigned to an individual group as the other three Ca II HK line indices, and neither is regarded as the most sensitive index in its FCA-clustered group. Moreover, when the efficiency to estimate metallicities of the K6 index for the whole sample is not clearly detected, the FCA tends to identify its dependence on other dominant physical parameters (e.g., the effective temperature). This may partly explain the case when K6 is categorized into the same group with Balmer line or $G$-band indices, which are usually sensitive to the effective temperature. To further testify the efficiency of such index exclusion method, we have carried out a test, where we randomly removed five line indices in corresponding groups as categorized in the case of $\lambda = 0.9965$, and reran the metallicity estimation through ANN. The results show that the precision of the FCA-ANN method will be degraded by a factor of ∼2 if line indices are removed in a random manner rather than the FCA-determined way.

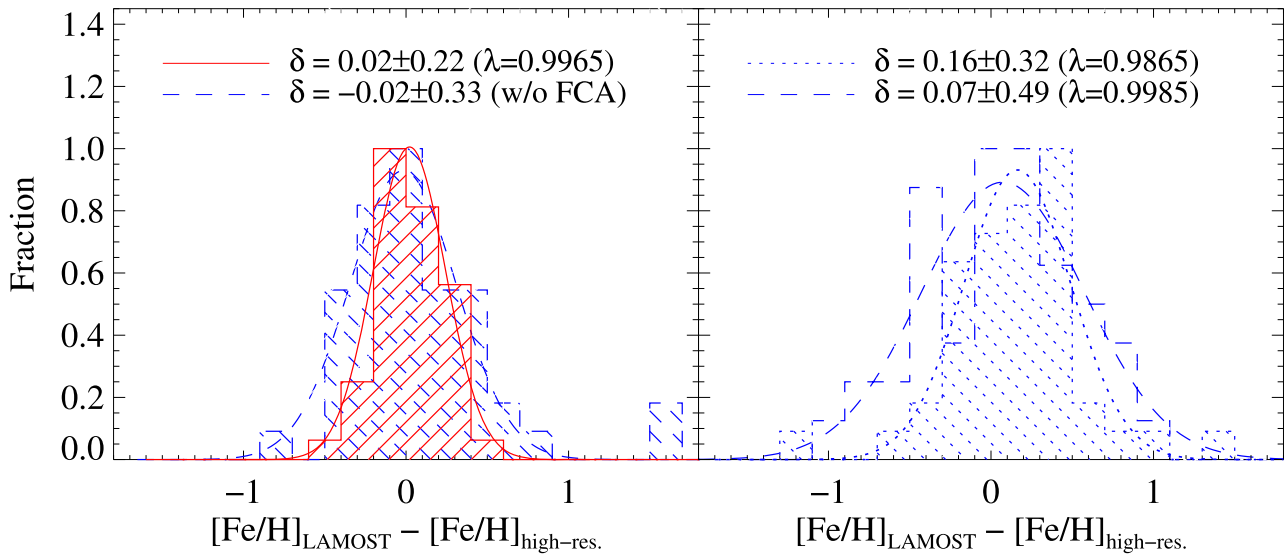

**Figure 3.** Left: Comparison of the metallicity uncertainties derived from FCA-ANN (threshold value $\lambda = 0.9965$, solid filled) and ANN (without FCA, dashed filled). Right: Comparison between results from FCA-ANN with threshold values of 0.9865 (dashed-dotted) and 0.9985 (dashed filled). Gaussian profiles have been fit to estimate the systematic offset and scatter. Note that only the test sample has been used to derive the distribution.

Nevertheless, the basic characteristic of fuzzy clustering is to divide the line indices into different categories, according to their similarity and equivalence relationship. Choosing a different threshold value (any $\lambda \in [0, 1]$) will result in a different way of grouping and thus a different number of groups, where the division between each group including the total number of groups is automatically defined once the threshold value is fixed. As an extreme example, there are two elements in the fuzzy similarity matrix, $r_{11} = 0.8566$ and $r_{21} = 0.8564$. When a threshold value of $\lambda = 0.9865$ has been adopted, the specific value of division between two neighbor groups is 0.8563, and $r_{11}$ and $r_{21}$ will be assigned into a same group; if the threshold value of grouping is $\lambda = 0.9965$, a specific value of division becomes 0.8565, and $r_{11}$ and $r_{21}$ will be assigned to two different groups (though the similarity coefficients are quite similar). Such a phenomenon is the result of normal operation in the process of FCA, and it usually reflects the actual situation. Otherwise, if two elements with significant numerical differences are grouped into the same category, it will inevitably lead to inaccurate predictions. To sum, line indices will be grouped differently if different threshold values are adopted, and the closer the threshold value to 1.0, the more groups the FCA will return. Therefore, when there is a sufficient number of groups, even indices with similar sensitivity could be categorized into different groups, as long as their specific values are not exactly the same (such as the cases adopting threshold values larger than 0.9965).

The left panel of Figure 3 compares the metallicity estimated for the test sample with (solid histogram) and without (dashed histogram) FCA clustering, where adopting the threshold value of 0.9965 has been used as an example. It is quite clear that after excluding five indices that have been identified as the least sensitive/relevant in their groups, reducing the input index dimension from 27 to 22, neither the accuracy or the dispersion of the resulted estimation of metallicities has been affected. We can also clearly see adopting different threshold values would lead to different results, as shown in the right panel of Figure 3. Moreover, such comparison based on our test sample indicates that there seems to be a preferred or optimized range for the choice of the threshold value $\lambda$. A $\lambda$ that either too large or too small would possibly lead to large uncertainties in the resulted estimation. Based on our tests on a number of different threshold values, it is recommended that a threshold value $\lambda = 0.9965$ or 0.9975 shall be the optimal choice for our sample/purpose. The preferred value is selected in such a way that the resulted systematic offset should be consistent with that directly derived before (without) adopting FCA clustering, and the offset and dispersion should not be significantly larger than that would be derived before adopting the FCA clustering. For example, in the case of a threshold value smaller than 0.9865 or larger than 0.9985, both the systematic offset and dispersion become larger, which indicates that the clustering does not provide sufficient information that is sensitive to the metallicity. And in the following discussion, the result obtained with a threshold value of 0.9965 will be adopted, which is considered to represent a preferred and optimized case for the FCA method on the VMP star sample studied here.

Note that as a test, this work only removes the least relevant line index for each group, those indices marked with "x" in the end in Table 2. Nevertheless, it is still quite clear that the FCA clustering is able to identify the relevance between line indices and the metallicity. For example in the case of 0.9965 (the 4th column), the discarded line indices include those that are not as sensitive to metallicities as other indices measuring similar features such as K6 (compared to K24 and K12), G1 (compared to G4300), or those that are measuring features that are not sensitive to metallicities themselves, such as Ca4227 and Fe4531, both of which correspond to metallic lines that are too weak for VMP stars in low-resolution spectra.

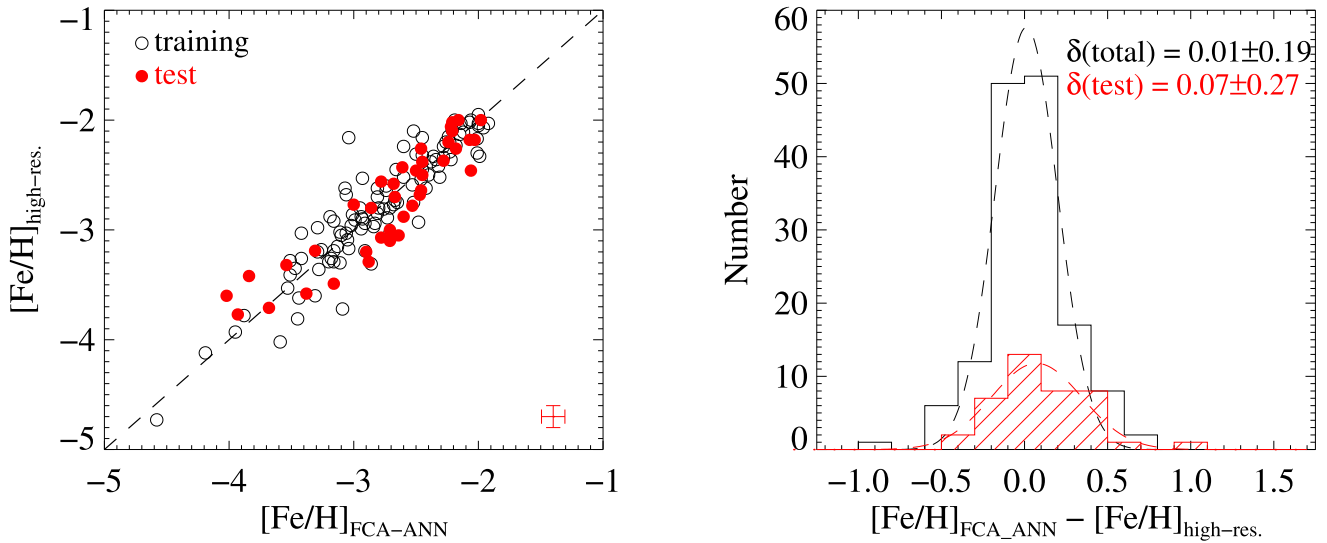

**Figure 4.** Left: Comparison between the FCA-ANN-derived metallicity and the high-resolution analysis. Results of the training sample and test sample are respectively shown in hollow and (red) filled circles. Right: Distribution of the measurement difference. The hollow and (dashed) filled histogram refer to the distribution of the whole sample (147 stars) and the test sample (44 stars), respectively, with the fitted Gaussian profiles shown in dashed lines to estimate the measurement error.

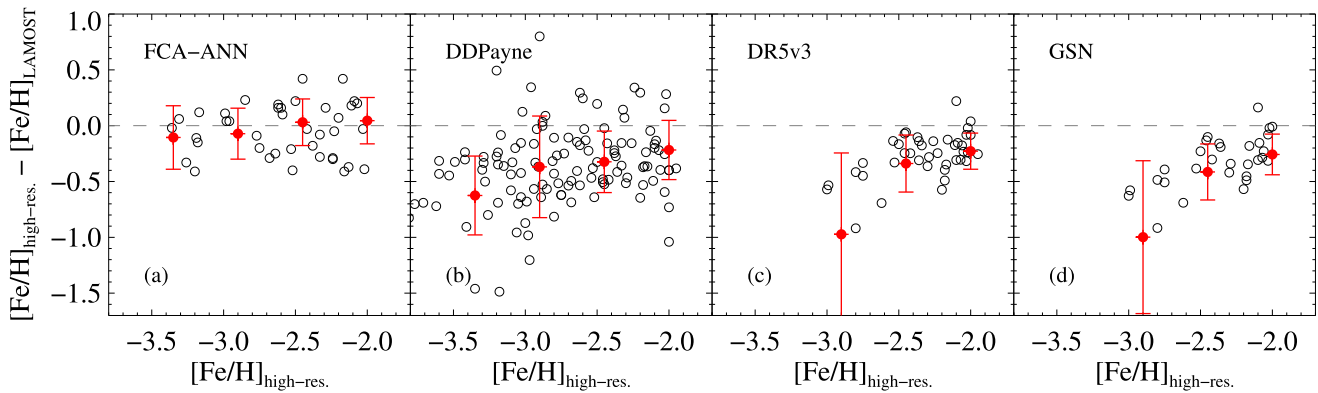

**Figure 5.** Distribution of the metallicity difference between high-resolution estimates and LAMOST low-resolution estimates, as a function of the metallicity derived from high-resolution spectra. Filled circles and error bars refer to the average difference and corresponding standard deviation in four ([Fe/H]).

### 3.2. Validation of the Method

Comparisons of the metallicity [Fe/H] derived from LAMOST spectra using the FCA-ANN method and those from high-resolution spectra are presented in Figure 4. As is shown, for the majority of the sample, the FCA-ANN analysis can derive rather consistent [Fe/H] with the high-resolution analysis, and a Gaussian fit to the metallicity difference indicates small offset of 0.01 dex and measurement uncertainty of 0.19 dex (for the whole sample). And even for the 44 test sample stars, the derived metallicities are still rather robust, with the offset and uncertainty, respectively, of 0.07 dex and 0.27 dex (larger scatter partly caused by the small number of the test sample).

Rather good consistency can even be found down to the extremely low-metallicity region with [Fe/H] < −3.0 However, it is also noticed that for a few targets, the FCA-ANN-derived metallicity shows relatively large deviation. After checking the targets and corresponding LAMOST spectra, these deviated objects usually turn out to be low-metallicity stars with higher temperatures whose absorption lines are so weak that the line indices measured in low-resolution spectra are not sufficiently sensitive enough to the metallicity, or objects with strong CH *G*-band (e.g., cool giants or carbon-enhanced metal-poor stars that could be prevalent at [Fe/H] < −2.5), which might bring certain bias to metallicity estimation.

Nevertheless in general, the FCA-ANN is capable to derive rather robust estimation of metallicities based on low-resolution spectra even for extremely and ultra metal-poor stars. Moreover, it should be noticed that although the tested sample of VMP stars cover quite wide range in $T_{eff}$ (as shown in Figure 1), and we have NOT done any preselection or classification on the temperature before estimating the metallicity. Therefore, involvement of line indices that covers spectral features more sensitive to physical parameters other than the metallicity (e.g., HD12, HD24), is able to help degenerate information from physical different parameters, and to derive reliable metallicities for objects with different spectral types.

It is also worth exploring whether the FCA-ANN-derived metallicity is sensitive to different metallicity range or spectra quality. Panel (a) of Figures 5 and 6 respectively reflect the behavior of the measurement uncertainty of [Fe/H] along different spectral quality (in the form of SNR at *g*-band), as well as the stellar metallicity (regarding the high-resolution result). It is clear that the FCA-ANN method is quite stable. Unlike the previous classical method, e.g., the template-matching method used by Li et al. (2018b), the FCA-ANN method does not strongly rely on the spectral quality, and it works just as fine on spectra with modest quality as for those with higher SNR. As shown in the panel (a) of Figures 5 and 6, there is no systematic difference between the FCA-ANN metallicity and the high-resolution results, and further more, the average scatter among different SNR ranges does differ a lot, all around $0.2 \sim 0.3$ dex, which is quite robust estimation for low-resolution spectral analysis. Besides, the uncertainty of metallicity for extremely metal-poor stars is not significantly larger than that for objects with "higher" metallicities, which means that it is a very reliable method for studies focusing on very to extremely metal-poor stars.

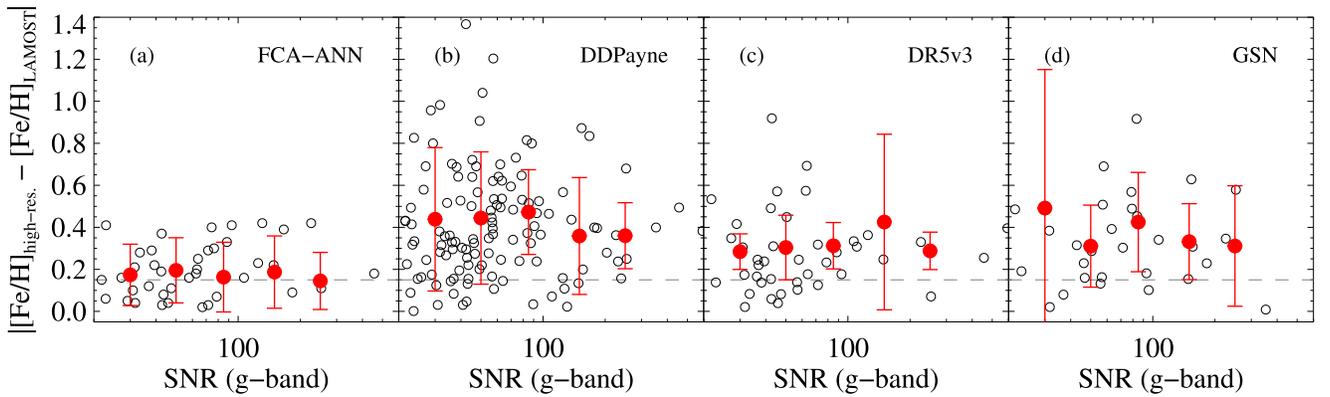

**Figure 6.** Distribution of the metallicity difference between high-resolution estimates and LAMOST low-resolution estimates, as a function of the SNR in the *g* band of LAMOST spectra. Filled circles and error bars refer to the average difference and corresponding standard deviation in five SNR bins).

### 3.3. Comparison with other Pipeline and Methods

Quite a number of efforts have been devoted to estimate stellar parameters from LAMOST spectra, e.g., the official LAMOST pipeline LASP, DDPayne, involving various methods. For example, LASP uses the UlySS package (Koleva et al. 2009) to derive the main stellar parameters including $T_{\rm eff}$, $\log g$, and [Fe/H]; Wang et al. (2019) derives the three stellar parameters as well as the average abundances of [$\alpha$/Fe] through generative spectrum networks (GSNs); DDPayne adopts deep machine-learning skills to derive stellar parameters together with abundances for a number of elements (Xiang et al. 2019). Although these methods obtain systematically consistent results compared with other studies, including high-resolution analysis, none of these are specifically designed to derive reliable metallicities for VMP stars, which is partially due to the limited parameter coverage in the metallicity space for the training sample or model templates, and therefore usually present systematic deviation from high-resolution analysis, as shown with the dashed histograms in Figure 7. Note that only program stars in our test sample have been considered for the following comparisons. By comparing the derived metallicity for common stars in the test sample with different methods, it is quite clear that the FCA-ANN could derive reliable metallicity for VMP stars. Furthermore, based on previous discussion, we believe that the FCA processed line indices should also be able to use as input for other methods.

Figure 5 compares the behavior of different methods for different metallicity ranges. It is shown that, despite of the systematic deviation of the other three methods, the typical uncertainty (scatter) is not significantly different at the region with [Fe/H] > −2.5; however, when it comes to extremely metal-poor region with [Fe/H] < −2.5, they are either unable to derive metallicities due to limitations of the adopted template grid (e.g., LASP and GSN), or present significant scatter as large as 0.5 dex (e.g., DDPayne). Moreover, we have also investigated the reliability of different methods in cases of different spectral quality. As can be seen in Figure 6, both FCA-ANN (panel (a)) and LASP (panel (c)) present stable measurement uncertainty along different SNR ranges, while the uncertainty of DDPayne (panel (b)) and GSN (panel (d)) clearly depends on the spectral quality, deriving more reliable metallicities for VMP stars with higher SNR.

### 4. Summary

This work has, for the first time, applied the mathematical method of fuzzy cluster analysis (FCA) to studying the physical parameters of stars. Specifically, we have tested it as a new method to group the observed line indices and reduce the dimension based on their relevance/sensitivity to the target parameter, in this case the stellar metallicity. For this early test, 27 line indices have been measured from the LAMOST low-resolution spectra for 147 very metal-poor (VMP) stars that have been analyzed with high-resolution spectroscopy. The measured line indices have then been clustered using the FCA method, and later used as input for the traditional artificial neural network (ANN) method to derive metallicities for these VMP stars. The test presented here show that the FCA method is able to automatically categorize the measured line indices according to their relevance/sensitivity to the metallicity. A number of different threshold values that controls the automatic FCA grouping of line indices have been adopted, leading to different results of categorization (including different number of groups, different assignment of line index to corresponding groups, etc.). Mathematically, the threshold value could be any number between 0 and 1, where the larger the value, the more groups will be resulted from the clustering, which could then leave less freedom for further dimension reduction of the input, and thus there would be a practical range for the choice of threshold value $\lambda$ for any specific case. Considering this test, in general, adopting any specific $\lambda$ between 0.9865 and 0.9990, the FCA method is able to reasonably identify line indices that are similarly/comparably sensitive to the stellar metallicity, and to determine the least relevant/sensitive line index in each group that shall be excluded for the next step.

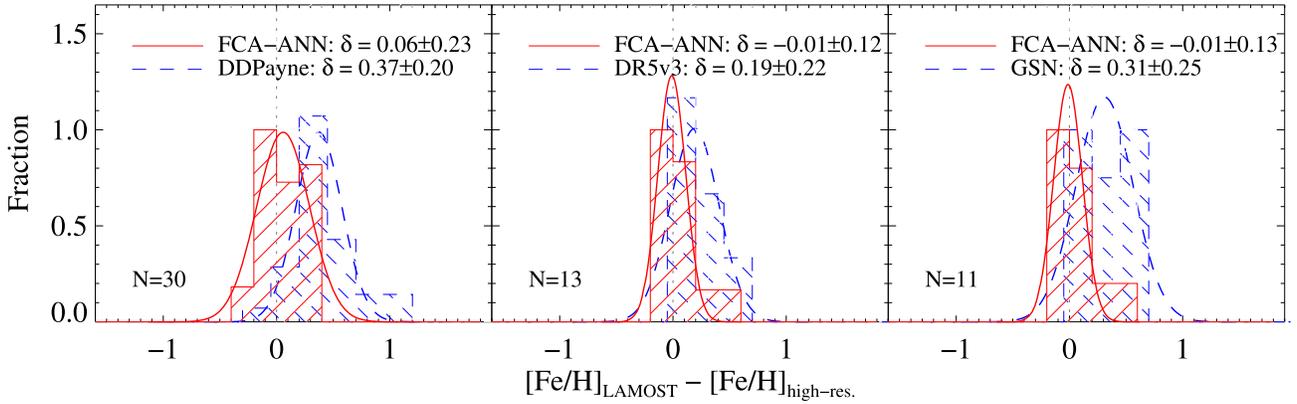

**Figure 7.** Comparison between the metallicity measured from LAMOST low-resolution spectra through the FCA-ANN method (red solid) and three other pipelines (blue dashed) including DDPayne (left), LASP (middle), and GSN (right), showing in the form of difference between high-resolution and LAMOST estimates. Systematic offsets and uncertainties are presented through Gaussian fitting, and shown in the plot. Note that all comparisons are based on the test sample of the FCA-ANN method, and the number of compared common stars are shown in each panel.

Based on our test, the optimal range of the threshold value for metallicity determination for this VMP star sample is $0.9965 \leqslant \lambda \leqslant 0.9975$, which could assign the 27 measured line indices into 10 groups and automatically exclude five indices before the next step. Comparison with the high-resolution results indicates that the specific (ANN) training on VMP stars could derive rather reliable metallicities for low-metallicity stars, and in general, after reducing the dimension of input line indices through FCA clustering, the estimation of metallicities still remain robust and do not lose any precision after dimension reduction, e.g., when adopting a preferred threshold value $\lambda = 0.9965$ or 0.9975, the offset and uncertainty of the derived metallicity are $-0.02$ dex and 0.33 dex, and 0.02 dex and 0.22 dex, respectively, before and after adopting FCA. The metallicity derived from LAMOST DR5 spectra through the FCA-ANN method has also been compared with that derived from other methods, including LASP, DDPayne, and GSN. It is shown that the performance of the FCA-ANN method is quite stable for different spectral quality from the metallicity of $[Fe/H] \sim -1.8$ down to $[Fe/H] < -3.5$, which thus provides an independent estimation on metallicities.

Compared with requiring precise numbers and assuming precise relations in classical clustering methods such as the K-means method, fuzziness and imprecision of the real observational data are considered during analysis in the FCA method, which is one obvious advantage of soft classification algorithms. The FCA method provides more flexible and applicable conditions for processing more information, e.g., in the case of this study, line indices or spectral data include degenerated information from multiparameters, including temperature, metallicity, surface gravity, etc. Therefore, compared with the traditional clustering methods, fuzzy classification through FCA can fully consider the fuzziness of the observed data, and is more suitable and obviously worth trying to be applied to astronomical data analysis. For this work, the FCA processed line indices have been transferred to the traditional ANN method as the input to derive the target parameter, metallicity, and it should NOT be limited to this test. We are still in the preliminary stage to explore how to apply the FCA method to the astronomical data, and both the size of the data used in this work as well as our experiences are too limited to fully understand the relation or reflection between math and astrophysics, The main purpose of this pilot project is to testify the feasibility of application of FCA to analyzing spectral data, which has been proved positive. Future explorations will be carried out for, e.g., tests with larger samples and on other physical parameters, combination with other methods, etc. And for future works, the FCA method shall also be used to cluster other observational measurements, to process the input for other methods (e.g., template matching, polynomial fitting) that could be used to determine physical parameters for stars.

Ongoing and future large scale low-resolution spectroscopic and narrow-band photometric survey projects, such as LAMOST-II, SDSS-V, Subaru/PFS, J-Plus/S-Plus, and the 2 m Chinese Space Survey Telescope survey, will provide an unprecedentedly huge database for studying various types and populations of stars in the Milky Way and the local universe. Working in the high-dimension database urgently requires efficient methods to identify the most relevant information and to reduce the dimension of the complex. Therefore, application of clustering methods such as FCA, which is able to consider the fuzziness and noncontinuity of the observational data, can be very helpful to effectively categorize and extract the most relevant/important information from the measurement, in the field of stellar physics, e.g., classification of stars, determination of physical parameters, identification of stars that present special features in the stellar spectra (and/or photometric colors) such as VMP stars and stars with abnormal elemental abundances.

The author would like to appreciate the anonymous referee for really helpful comments and suggestions. This work is supported by the NSFC grants No. 11988101 and 11973049, National Key R&D Program of China No. 2019YFA0405502, the Strategic Priority Research Program of Chinese Academy of Sciences, grant No. XDB34020205, and the Youth Innovation Promotion Association of Chinese Academy of Science (No. Y202017). We acknowledge the science research grants from the China Manned Space Project with No. CMS-CSST-2021-B05. Guoshoujing Telescope (the Large Sky Area Multi-Object Fiber Spectroscopic Telescope, LAMOST) is a National Major Scientific Project built by the Chinese Academy of Sciences. Funding for the project has been provided by the National Development and Reform

Commission. LAMOST is operated and managed by the National Astronomical Observatories, Chinese Academy of Sciences.

*Facility:* LAMOST.

*Software:* MATLAB.

## Appendix A
## Basic Definition and Theorems of the Fuzzy Theory

We here briefly present the *basic definition and theorems* of the fuzzy theory, starting from a number of key definitions.

**Definition 1**–Fuzzy mathematics. Fuzzy mathematics is a set of mathematical methods to describe sets where the membership of elements is not binary, and here, fuzzy means ambiguity. It is a new mathematical subject developed after classical mathematics and statistical mathematics (see Zadeh 1965 for references). The word "fuzzy" itself means "vague." Throughout this paper, the word "fuzzy" corresponds to special mathematical symbols, formulas, and methods related to the theory of fuzzy mathematics.

**Definition 2**–Fuzzy clustering. Fuzzy clustering is a mathematical method that describes and classifies things by using methods and tools from fuzzy mathematics. The basic rule of fuzzy clustering is to divide the data set into several classes or clusters, in accordance with the principle of "minimize the similarity between classes and maximize the similarity within classes"; the differences between samples from different classes should be as large as possible, while the differences between samples within each class should be as small as possible. The data are classified by similarity, as determined by a fuzzy similarity relation. Through fuzzy clustering, each sample will be assigned to a category with a certain degree of membership/probability. The uncertainty degree of belonging to the category will be derived during fuzzy clustering, which thus more accurately reflects the actual situation.

**Definition 3**–Fuzzy matrix. A matrix in which all the elements have values within the closed interval of [0,1] is called a fuzzy matrix.

**Definition 4**–Hard classification. A sample is definitely assigned to one of the classes. Most traditional classification methods, including the commonly used K-means clustering, belong to hard classification.

**Definition 5**–Soft classification. The probability that a sample belongs to each class corresponds to a membership vector and the classification result is the one with the highest probability, where the sum of all elements in the vector is 1. The FCA is a typical soft classification method.

**Definition 6**–Membership function. We use $\mu_{\tilde{A}(x)}$ to describe the degree to which $x$ belongs to the fuzzy set $\tilde{A}$, which is called the membership function of the fuzzy set $\tilde{A}$. The closer $\mu_{\tilde{A}(x)}$ is to 1, the more likely $x$ belongs to $\tilde{A}$. Suppose $\Omega$ refers to our sample space composed of measured line indices, and $x$ is an element of $\Omega$. If there is a number $\tilde{A}(x)$ is in the interval [0,1] corresponding to any element $x$ of $\Omega$, $\tilde{A}$ is called a fuzzy set over $\Omega$. Here, the mathematical expression of the fuzzy set is:

$$\tilde{A} = \{(x, \mu_{\tilde{A}}(x)) | x \in \Omega\}, \quad (A1)$$

where $\Omega$ is the sample space composed of measured data.

**Definition 7**–Square method. In fuzzy clustering analysis, to transform similar matrices into equivalent matrices, successive square multiplication of similar matrices is needed in the calculation process. In fact, every calculation is a matrix square multiplication operation until the transitive closure is determined. The transitive closure of similarity relation $R$ is calculated based on the following expression:

$$\tilde{R} \to \tilde{R}^2 \to \tilde{R}^2 \to \cdots \to \tilde{R}^{2^k} = \tilde{R}^{2^{k+1}}. \quad (A2)$$

Because it is a successive square, this algorithm is called the "square method"

**Definition 8**–Reflexivity, symmetry, and transivity. Suppose $\Omega$ is a finite set, a fuzzy relation $\tilde{R}$ exists on $\Omega$, and its corresponding fuzzy matrix is therefore $\tilde{R} = (r_{ij})_{n \times m}$. If reflexivity, symmetry, and transitivity are satisfied, then $\tilde{R}$ is regarded to be a fuzzy equivalent matrix, and its relation is fuzzy equivalent relation. If the fuzzy matrix only satisfies reflexivity and symmetry, it is considered a similarity relation, where the criteria for reflexivity, symmetry, and transivity are defined as follows:

$$\text{Reflexivity: } r_{il} = 1 (i = 1, 2, \ldots, n), \quad (A3)$$

$$\text{Symmetry: } r_{ij} = r_{ji} = 1, (i, j = 1, 2, \ldots, n), \quad (A4)$$

$$\text{Transitivity: } \tilde{R} \circ \tilde{R} \subseteq \tilde{R}. \quad (A5)$$

The fuzzy cluster analysis in this paper is thus according to the above definitions.

## Appendix B
## The VMP Star Target List

Here we list all the 147 VMP stars that have been observed with high-resolution spectra and used for the study in this work. Metallicities measured from LAMOST spectra through the FCA-ANN method are presented in columns 8 through 15, including eight different choices of the threshold value. The 27 used line indices are shown in columns 16 through 42. Those measured from LAMOST spectra by other pipelines are presented in columns 43, 44, and 45, respectively, for DDPayne, DR5v3, and GSN.

References of the literature high-resolution analyses are also presented in column 46.

The high-resolution spectroscopic analysis results used in this work include the following works: Aoki et al. (2002), Aoki et al. (2005), Aoki et al. (2007), Aoki et al. (2013), Aoki et al. (2017), Aoki et al. (2018), Allen et al. (2012), Andrievsky et al. (2009), Barklem et al. (2005), Boesgaard (2007), Boesgaard et al. (2011), Bonifacio et al. (2009), Bonifacio et al. (2018), Carney et al. (2003), Charbonnel & Primas (2005), Christlieb et al. (2004), Cohen et al. (2006), Cohen et al. (2013), Caffau et al. (2005), Caffau et al. (2011), Caffau et al. (2018), Frebel et al. (2007), Goswami et al. (2006), Hansen et al. (2012), Hansen et al. (2015), Hansen et al. (2018), Hollek et al. (2011), Holmbeck et al. (2018), Honda et al. (2004), Ishigaki et al. (2013), Jacobson et al. (2015), Lai et al. (2007), Lai et al. (2008), Li et al. (2015a), Li et al. (2015b), Li et al. (2015c), Li et al. (2018a), Masseron et al. (2012), Matsuno et al. (2017), Meléndez et al. (2010), Nissen et al. (2007), Placco et al. (2016), Preston & Sneden (2000), Rich & Boesgaard (2009), Roederer et al. (2014), Ruchti et al. (2011), Sakari et al. (2018), Schlaufman & Casey (2014), Schuler et al. (2007), Spite et al. (2013), Spite et al. (2018), Susmitha Rani et al. (2016), Venn et al. (2004), Yong et al. (2013).

## Appendix C
## Empirical Correction on the Metallicity in the Low-Metallicity Region

Based on the comparison in Section 2.4, it is clear that the three pipelines that have been discussed for comparison are not specifically designed to cover the very low-metallicity range. In general, there exist systematic offsets between the pipeline-estimated metallicity and the high-resolution derived value (e.g., the distribution in Figure 7). Therefore, we try to provide an empirical correction on the low-resolution estimated [Fe/H] from these data sets based on comparisons with the targets listed in Table 3, the high-resolution analyzed sample that has been used for this paper. As shown in Figure 8, linear fittings are generally sufficient for the correction. Note that, since the DR5v3 and GSN data sets only include a few dozens of VMP stars (whose low-metallicity range only reaches to [Fe/H] ∼ −2.5), and are quite agreeable with each other concerning the derived metallicity for metal-poor stars, these two sets of data are combined to derive the correction, the right panel in Figure 8 (with circles referring to the results from DR5v3, and triangles referring to those from GSN).

The derived correction on the metallicity is shown in Figure 8, and it can be seen that all the three pipelines tend to overestimate the metallicity for VMP stars, and the slopes of the fitted trend are quite similar, which is probably due to the fact that the training samples of these pipelines are dominated by relatively metal-rich stars, e.g., the APOGEE data. However, it seems that if adopting the derived correction, one could still use the pipeline-estimated metallicities for statistical purposes in studying metal-poor stars.

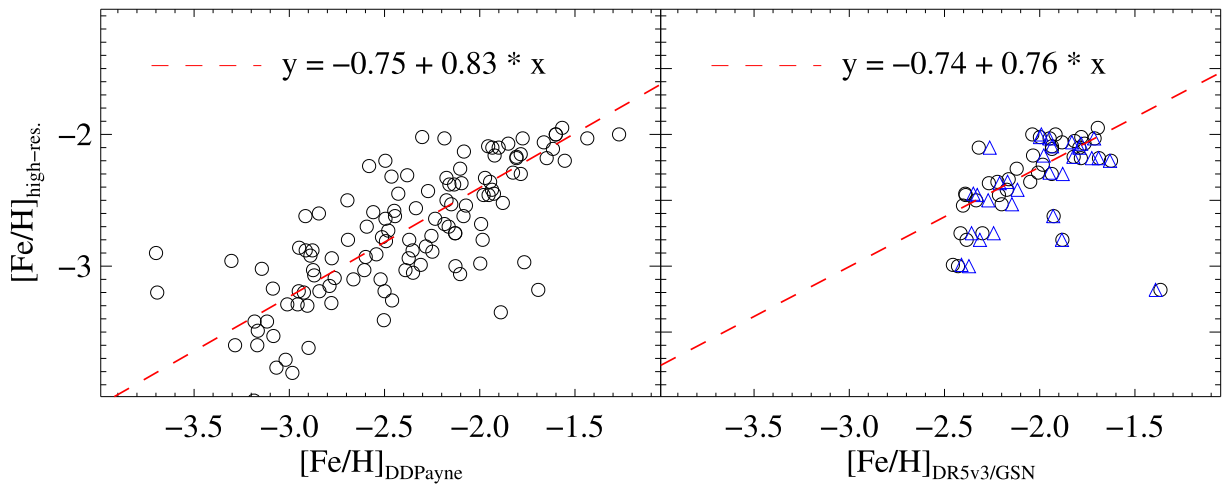

**Figure 8.** Comparison between high-resolution derived metallicities and LAMOST low-resolution estimates, based on the results from DDPayne (left) and DR5v3 (right, circles) and GSN (right, triangles). Linear fittings have been derived as a toy correction for all these data sets, as shown in dashed lines.

**Table 3**
List of the Used VMP Stars

| Column (1) | (2) | Unit (3) | Description (4) |
|---|---|---|---|
| 1 | R.A. | deg | R.A. |
| 2 | decl. | deg | decl. |
| 3 | $T_{\rm eff}$_hires | K | Effective temperature derived from high-resolution spectra |
| 4 | $\log g$_hires |  | Surface gravity derived from high-resolution spectra |
| 5 | [Fe/H]_hires |  | Metallicity derived from high-resolution spectra |
| 6 | sn_g |  | SNR at $g$-band of the LAMOST spectra |
| 7 | sn_r |  | SNR at $r$-band of the LAMOST spectra |
| 8 | [Fe/H]_FCA865 |  | Metallicity derived from LAMOST spectra, FCA clustered, $\lambda = 0.9865$ |
| 9 | [Fe/H]_FCA885 |  | Metallicity derived from LAMOST spectra, FCA clustered, $\lambda = 0.9885$ |
| 10 | [Fe/H]_FCA965 |  | Metallicity derived from LAMOST spectra, FCA clustered, $\lambda = 0.9965$ |
| 11 | [Fe/H]_FCA975 |  | Metallicity derived from LAMOST spectra, FCA clustered, $\lambda = 0.9975$ |
| 12 | [Fe/H]_FCA980 |  | Metallicity derived from LAMOST spectra, FCA clustered, $\lambda = 0.9980$ |
| 13 | [Fe/H]_FCA983 |  | Metallicity derived from LAMOST spectra, FCA clustered, $\lambda = 0.9983$ |
| 14 | [Fe/H]_FCA985 |  | Metallicity derived from LAMOST spectra, FCA clustered, $\lambda = 0.9985$ |
| 15 | [Fe/H]_FCA990 |  | Metallicity derived from LAMOST spectra, FCA clustered, $\lambda = 0.9990$ |
| 16 | CaHK | Å | Line index CaHK |
| 17 | K24 | Å | Line index K24 |
| 18 | K12 | Å | Line index K12 |
| 19 | K6 | Å | Line index K6 |
| 20 | He I4026 | Å | Line index He I4026 |
| 21 | HD24 | Å | Line index HD24 |
| 22 | HD12 | Å | Line index HD12 |
| 23 | CN1 | Å | Line index CN1 |
| 24 | CN2 | Å | Line index CN2 |
| 25 | Ca4227 | Å | Line index Ca4227 |
| 26 | G4300 | Å | Line index G4300 |
| 27 | G1 | Å | Line index G1 |
| 28 | Fe4383 | Å | Line index Fe4383 |
| 29 | Ca4455 | Å | Line index Ca4455 |
| 30 | He I4471 | Å | Line index He I4471 |
| 31 | Fe4531 | Å | Line index Fe4531 |
| 32 | Hbeta | Å | Line index Hbeta |
| 33 | Fe5015 | Å | Line index Fe5015 |
| 34 | Mg1 | Å | Line index MG1 |
| 35 | Mg2 | Å | Line index Mg2 |
| 36 | Mgb | Å | Line index Mgb |
| 37 | Fe5270 | Å | Line index Fe5270 |
| 38 | Fe5335 | Å | Line index Fe5335 |
| 39 | Fe5406 | Å | Line index Fe5406 |
| 40 | Fe5709 | Å | Line index Fe5709 |
| 41 | Fe5782 | Å | Line index Fe5782 |
| 42 | NaD | Å | Line index NaD |
| 43 | [Fe/H]_DDPayne |  | Metallicity derived from LAMOST spectra, from DDPayne |
| 44 | [Fe/H]_DR5v3 |  | Metallicity derived from LAMOST spectra, from official DR5v3 |
| 45 | [Fe/H]_GSN |  | Metallicity derived from LAMOST spectra, from GSN |
| 46 | ref_hires |  | Reference of high-resolution analyses |

**Note.** The table is published in its entirety in the machine-readable format. A portion is shown here for guidance regarding its form and content.

(This table is available in machine-readable form.)

## ORCID iDs

Haining Li  https://orcid.org/0000-0002-0389-9264